\documentclass[letterpaper, 10 pt, conference]{ieeeconf}  

\IEEEoverridecommandlockouts                    

\usepackage[utf8]{inputenc}
\usepackage{lipsum}
\usepackage{graphicx}
\usepackage{amsmath}
\usepackage{amssymb}
\usepackage{indentfirst}
\usepackage[english]{babel}
\usepackage{multirow}
\usepackage{tabularx}
\usepackage{cases}
\usepackage{dirtytalk}
\usepackage{caption}
\usepackage{multicol}
\usepackage[font={small}]{caption}
\usepackage{cite}
\usepackage{makecell}
\usepackage{algorithm}
\usepackage{algpseudocode}

\usepackage{xcolor}
\providecommand{\keywords}[1]
{
  \small	
  \textbf{\textit{Keywords---}} #1
}

\newtheorem{theorem}{Theorem}
\newtheorem{problem}{Problem}
\newtheorem{lemma}{Lemma}

\newtheorem{definition}{Definition}
\newtheorem{remark}{Remark}

\begin{document}

\title{Plug-in Estimation Method for Topology and Admittance Estimation in Electric Grids}

\author{Norak Rin, I. Shames, Ian R. Petersen, Nanduni I. Nimalsiri, and Elizabeth L. Ratnam
 \thanks {N. Rin and I. R. Petersen are with the Research School of Engineering, The Australian National University, Canberra, Australia. (E-mail: norak.rin@anu.edu.au)} 
 \thanks{I. Shames is with the Department of Electrical and Electronic Engineering, The University of Melbourne, Victoria, Australia. }
 \thanks{N. Rin, Nanduni I. Nimalsiri, and E. L. Ratnam are with the Energy Research Unit, CSIRO, Australia.}}

\maketitle
\begin{abstract}
This paper proposes a scalable method for topology and admittance estimation in electric power grids. Our proposed plug-in estimation method incorporates precise phasor measurement unit data to estimate planned or unplanned grid topology and admittance parameter changes. Our plug-in estimation method decouples the problem into two stages: (i) constructing and averaging voltage coefficient matrices and current vectors from phasor data, and (ii) estimating admittance parameters via a least-squares regression approach with thresholding to recover the network topology. Rigidity matrix theory underpins the design of our plug-in estimation method --- to determine the data requirements (e.g., necessary and sufficient conditions) for accurate estimation of the grid topology and admittance parameters. By means of numerical simulations on the IEEE 13-node, 123-node, and European low voltage test feeders, we demonstrate the scalability of our plug-in estimation method and benchmark the computational time against a structured total least squares method. Our simulation results highlight the practicality of the proposed method for topology and admittance estimation in large-scale electric networks, including electric grids with partial topology information.
\end{abstract}

\section{Introduction}

The recent grid integration of distributed and renewable energy resources is transforming the operation of modern electric power systems \cite{Admittance}. Concurrently, increasing deployments of phasor measurement units (PMUs) have enabled the collection of high-resolution voltage and current phasor data (i.e., synchrophasors) at increasing spatial and temporal resolution \cite{Precision}. To support the energy transition to renewables, synchrophasor data provides new opportunities to estimate the grid topology and admittance parameters under evolving grid conditions \cite{Learning}.

Topology identification of the electric grid supports critical grid operations --- including state estimation, power outage restoration, and distributed energy resources coordination. Real-time topology identification methods can provide an accurate operational model instead of outdated switch-state single-line diagrams~\cite{Learning}. Jointly identifying grid topology and estimating line admittance can enhance operations under evolving grid conditions, that is, by enabling timely recalibration of grid controllers and impedance protection settings to mitigate misoperation risks.~\cite{8122027}

Many authors have proposed regression-based methods for electric grid admittance estimation assuming knowledge of the grid topology. For example, Lang et al.~\cite{Structure} employ a structured total least squares method to estimate admittance parameters of a power distribution network. Mishra and de Callafon in \cite{9930858} propose a least squares approach to estimate three-phase line admittance parameters in either a transmission or distribution grid using synchrophasor measurements. Similarly, Gupta et al.~\cite{9464332} propose a linear estimation model to accurately estimate the admittance matrix of a three-phase distribution grid using noisy synchrophasor measurements.

Several authors have also proposed regression-based estimation methods, underpinned by the \emph{plug-in} principle, that are scalable in the presence of noisy data~\cite{wright2011using}. That is, the plug-in principle is a technique commonly used in statistics to estimate features in a distribution of data~\cite{cuevas1997plug}. For example, the authors in \cite{wright2011using} propose the plug-in principle to evaluate unknown regression parameters by substituting empirical sample data into a theoretical model. In signal processing applications involving large-dimensional data, the authors in~\cite{wang2017extreme} propose a plug-in estimation method to approximate unknown noise covariance matrices for optimal state estimation. However, there are limited applications of the plug-in principle to estimate admittance parameters in the absence of grid topology information that are scalable to both transmission and distribution estimation problems.

In this paper, we propose an estimation method for electric grid topology and admittance parameters by integrating graph rigidity theory with the plug-in principle. Our approach decomposes the estimation problem into two stages: (i) the construction and averaging of voltage coefficient matrices and current vectors from time-series phasor data, and (ii) the estimation of admittance parameters through a least-squares regression followed by a thresholding step to identify non-existent edges. That is, the plug-in estimation method avoids the need for large matrix inversion, significantly reducing the computational burden. Accordingly, our estimation method is scalable to electric distribution grids with hundreds of nodes and is robust against measurement noise.

The organization of this paper is as follows. In Section~{\ref{sect_preliminaries}}, we define the system of equations for an electric grid based on Kirchhoff's laws. In Section~{\ref{Section3}}, we propose a plug-in estimation method integrated with graph rigidity theory to estimate the grid topology and admittance parameters. In Section~{\ref{section4}}, we present our numerical simulations, which is followed by the Conclusion.

\subsection*{Notation}

Let $\mathbb{R}^n$ and $\mathbb{C}^n$ denote the sets of $n$-dimensional vectors of real and complex numbers, respectively. Let $\mathbb{S}^n$ denote the set of $n\times n$ complex symmetric matrices. For a matrix $\textbf{\textit{X}}\in\mathbb{R}^{m\times n}$ (or $\mathbb{C}^{m\times n}$) with $m$-by-$n$ dimensions, $X_{i,j}$ denotes the element corresponding to row $i$ and column $j$, and $\textbf{\textit{X}}^\top\in\mathbb{R}^{n\times m}$ (or $\mathbb{C}^{m\times n}$) denotes its transpose matrix.
Let $\mathbf{1}_{n}\in\mathbb{R}^{n}$ and $\mathbf{0}_{n}\in\mathbb{R}^{n}$ denote $n$-dimensional vectors which have all elements equal to $1$ and $0$, respectively. Let $\textbf{\textit{D}}=\mathrm{diag}\left(a_1,a_2,\dotsc,a_n\right)$ $\in\mathbb{R}^{n\times n}$ (or $\mathbb{C}^{n\times n}$) denote a diagonal matrix where
\begin{equation}\nonumber
 D_{i,j} = 
    \begin{cases}
      a_i & \text{if $i = j \in\{1,2,\dotsc,n\}$},\\
      0 & \text{if $i \neq j$.}
    \end{cases}       
\end{equation}
Throughout, boldface represents vector or matrix quantities whereas scalar quantities are non-bold.


\section{Preliminaries}\label{sect_preliminaries}

We introduce notation and concepts supporting our approach to estimate the topology and admittance parameters of an electric grid. Our approach assumes no prior information, is underpinned by graph theory, and can be applied to any kind of grid topology (e.g., radial, meshed, unbalanced, etc).

Let the electric network be modeled as a \emph{complete graph} $\mathcal{G}(\mathcal{N,E})$, where $\mathcal{N}= \{1,2,\dotsc,n\}$ is the set of nodes. Also, $\mathcal{E}= \{(i,j)| i\in\mathcal{N}\; ,j\in\mathcal{N},\; i<j\}$ is the set of edges,
where the edge $(i,j)\in \mathcal{E}$ connects node $i$ and node $j$ so that it leaves node $i$ and enters node $j$. Note that we use this ordered pair convention for ease of presentation, but the underlying graph is assumed to be undirected. The graph $\mathcal{G}(\mathcal{N,E})$ is also assumed to be a \emph{weighted graph} where each edge $(i,j)\in \mathcal{E}$ is assigned a weight $\alpha_{i,j}$. Let $e$ denote the cardinality of the set $\mathcal{E}$, such that $|\mathcal{E}| = e$. Then, by definition of a complete graph for $\mathcal{G}(\mathcal{N},\mathcal{E})$, $e = n(n-1)/2$. Let $\textbf{\textit{H}} \in \mathbb{R}^{n\times e}$ denote the \emph{incidence matrix} of the undirected subgraph $\mathcal{G}(\mathcal{N},\mathcal{E})$. The $il$-th element of $\textbf{\textit{H}}$, denoted by $H_{i,l}$, with $i \in \mathcal{N}$ and $l \in \mathcal{E}$, is defined by 
 \begin{equation}\label{equation1}
     H_{i,l} := 
    \begin{cases}
      1 & \text{if edge $l$ leaves node $i$},\\
      -1 & \text{if edge $l$ enters node $i$},\\
      0 & \text{otherwise}.
    \end{cases}
 \end{equation}

\begin{figure}[!t]
    \centering
\includegraphics[width=0.6\linewidth]{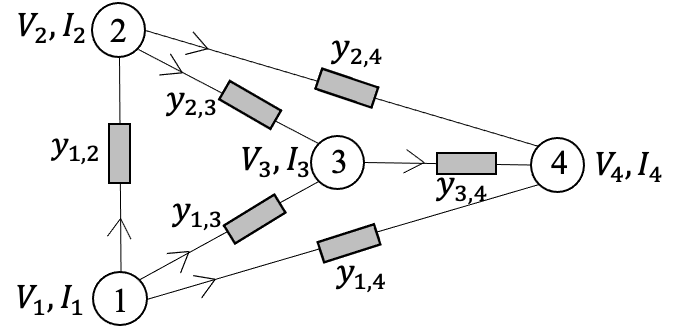}
    \caption{An example graph $\mathcal{G}(\mathcal{N,E})$ where $n=4$ and network admittance parameters are represented by $y_{1,2}, y_{1,3}, y_{1,4}$, $y_{2,3}, y_{2,4}, y_{3,4}$. } \label{Fig1}
\end{figure}

In what follows, we will use the graph $\mathcal{G}(\mathcal{N},\mathcal{E})$ to represent any complete network of admittance. Fig.~\ref{Fig1} illustrates an example of a complete graph $\mathcal{G}(\mathcal{N,E})$, where $n = 4$. Here, the weight of each edge $(i,j)\in \mathcal{E}$ corresponds to an \emph{admittance parameter}, $y_{i,j} \in \mathbb{C}$; e.g., the admittance of a transmission line or a transformer. The \emph{voltage} at node $j\in \mathcal{N}$ is denoted by $V_j \in\mathbb{C}$, and it is measured with respect to some ground potential. Similarly, the \emph{current} at node $j\in \mathcal{N}$ is denoted by $I_j \in\mathbb{C}$. Note that in this construction of the electric grid, we assume that there is no ground node in $\mathcal{N}$\footnote{Extensions to include a ground node are straightforward and result in increasing number of nodes and number of unknown admittance parameters.}.

 To develop the relationship between $V_j$ and $I_j$, where $j\in \mathcal{N}$, consider the graph $\mathcal{G}(\mathcal{N},\mathcal{E})$. For all nodes in $\mathcal{G}(\mathcal{N},\mathcal{E})$, the \emph{voltage vector} is denoted by $\textbf{\textit{V}} = [V_1,V_2,\dotsc, V_n]^\top \in \mathbb{C}^{n}$ and the \emph{current vector} is denoted by $\textbf{\textit{I}} = [I_1,I_2,\dotsc,I_n]^\top \in \mathbb{C}^{n}$. Accordingly, the \emph{admittance matrix} is denoted by $\textbf{\textit{Y}} \in \mathbb{S}^{n}$, where each element of $\textbf{\textit{Y}}$ is defined by
 \begin{equation*}
     Y_{i,j} := 
    \begin{cases}
      \sum\limits_{\substack{k = 1, \\ k\neq i}}^{n} {y_{i,k}} & \text{if $i=j$},\\
      -y_{i,j} & \text{if $i\neq j$}.
    \end{cases}
 \end{equation*}
According to Kirchhoff's laws \cite[Chapter 10]{Basic}, the admittance matrix  satisfies the relationship between $\textbf{\textit{V}}$ and $\textbf{\textit{I}}$ as 
 \begin{equation}\label{eq1}
 \textbf{\textit{I}} = \textbf{\textit{Y}}\textbf{\textit{V}},
 \end{equation}
 where $\textbf{\textit{Y}}\mathbf{1}_n = \mathbf{0}_n$. 
 
 Next, we rewrite the relationship between \textbf{\textit{V}} and \textbf{\textit{I}} by incorporating the incidence matrix \textbf{\textit{H}}. Corresponding to the complex symmetric matrix \textbf{\textit{Y}}, we define a complex \emph{admittance vector} by ${\textbf{\textit{y}}} := [-Y_{2,1}, -Y_{3,1}, \dotsc, -Y_{n,1}, -Y_{3,2}, -Y_{4,2}, \dotsc, -Y_{n,n-1}]^\top \in \mathbb{C}^e.$ Note that in the definition of \textbf{\textit{y}}, we use the fact that $\textbf{\textit{Y}}\mathbf{1}_n = \mathbf{0}_n$ to remove the redundant diagonal elements of \textbf{\textit{Y}}; see also \cite{9858017}. Using this notation, we can rewrite~\eqref{eq1} as
  \begin{equation}\label{eq2}
      \textbf{\textit{I}} = \textbf{\textit{H}} \mathrm{diag}(\textbf{\textit{H}}^\top \textbf{\textit{V}})\textbf{\textit{y}}.
  \end{equation}
  We denote by $\Tilde{\textbf{\textit{A}}}(\textbf{\textit{V}}) = \textbf{\textit{H}} \mathrm{diag}(\textbf{\textit{H}}^\top \textbf{\textit{V}})\in \mathbb{C}^{n\times e}$ 
  the \emph{voltage coefficient matrix}. Then,~\eqref{eq2} can be equivalently written as
 \begin{equation}\label{eq3}
 \textbf{\textit{I}} = \Tilde{\textbf{\textit{A}}}(\textbf{\textit{V}})\textbf{\textit{y}}.
 \end{equation}

In deriving~\eqref{eq3}, we considered a single measurement of the voltage vector \textbf{\textit{V}} and the current vector \textbf{\textit{I}}. Next, we consider a series of voltage and current measurements (at different operating points) \textbf{\textit{V}} and \textbf{\textit{I}} in order to uniquely identify the grid topology and admittance parameters. Suppose that different measurements of \textbf{\textit{V}} and \textbf{\textit{I}} are indexed by $k \in \mathcal{T}=\{1,2,\dotsc,T\}$. That is, $\textbf{\textit{V}}^{(k)} \in \mathbb{C}^n$ denotes the voltage vector measured corresponding to $k\in \mathcal{T}$ and $\textbf{\textit{I}}^{(k)} \in \mathbb{C}^n$ denotes the current vector measured corresponding to $k\in \mathcal{T}$. In the sequel, we assume that the pair vectors $(\textbf{\textit{V}}^{(k)}, \textbf{\textit{I}}^{(k)})$ are known exactly for all $k\in\mathcal{T}$ and that the measurements are such that each voltage vector $\textbf{\textit{V}}^{(1)},\textbf{\textit{V}}^{(2)},\dotsc,\textbf{\textit{V}}^{(T)}$ and each current vector
$\textbf{\textit{I}}^{(1)},\textbf{\textit{I}}^{(2)},\dotsc,\textbf{\textit{I}}^{(T)}$ are distinct from one another, that is, $V^{(k)}_j \neq V^{(\Tilde{k})}_j$ and $I^{(k)}_j \neq I^{(\Tilde{k})}_j$, where $j\in\mathcal{N}$, $k\in\mathcal{T}$, $\Tilde{k}\in\mathcal{T}$, and $k \neq \Tilde{k}$. We call the pair vectors $(\textbf{\textit{V}}^{(k)}, \textbf{\textit{I}}^{(k)})$ measured at $k\in\mathcal{T}$ an \emph{operating point/condition} in an electric power grid. It now follows from~\eqref{eq3} that 
 \begin{equation}\label{eq4}
     \textbf{\textit{I}}^{(k)} = \Tilde{\textbf{\textit{A}}}(\textbf{\textit{V}}^{(k)})\textbf{\textit{y}}, \forall k \in \mathcal{T},
 \end{equation}
 where $\Tilde{\textbf{\textit{A}}}(\textbf{\textit{V}}^{(k)})$ denotes the voltage coefficient matrix constructed from the voltage vector measured at $k\in \mathcal{T}$.
 
 In practice, measurements are corrupted by some noise. In this paper, we assume a timescale separation in grid operating conditions (e.g., feeder operating conditions change approximately every 15 minutes \cite{9276412}) against changes in voltages and currents corresponding to measurement noise (e.g., feeder load changes every sub-second \cite{7741972,brady2020phasor}). In what follows, we re-write equation~\eqref{eq4} for all operating conditions $k\in\mathcal{T}$ considering the presence of noisy measurements in both the input
currents and the measured voltages.

At each operating condition $(\textbf{\textit{V}}^{(k)}, \textbf{\textit{I}}^{(k)})$ measured at $k\in\mathcal{T}$, we consider time-series voltage and current measurements corrupted by some noise. Specifically, we collect $B$ time-series voltage measurements $\textbf{\textit{V}}^{(k)}_1,\textbf{\textit{V}}^{(k)}_2,\dotsc,\textbf{\textit{V}}^{(k)}_B$ and $B$ current measurements $\textbf{\textit{I}}^{(k)}_1,\textbf{\textit{I}}^{(k)}_2,\dotsc,\textbf{\textit{I}}^{(k)}_B$ at operating condition $k\in\mathcal{T}$. Throughout, we denote a \emph{batch} time-series voltage and current measurements corresponding to operating condition $k\in\mathcal{T}$ by the pair $(\textbf{\textit{V}}^{(k)}_t, \textbf{\textit{I}}^{(k)}_t)$ for all $t \in \{1,2,\dotsc,B\}$. Next, we compute batch averages, denoted by $\overline{\textbf{\textit{V}}}^{(k)} = \frac{1}{B}\sum_{t=1}^{B}\textbf{\textit{V}}^{(k)}_t$ and $\overline{\textbf{\textit{I}}}^{(k)} = \frac{1}{B}\sum_{t=1}^{B}\textbf{\textit{I}}^{(k)}_t$. Then, the voltage coefficient matrix corresponding to a batch of measurements is $\overline{\textbf{\textit{A}}}(\overline{\textbf{\textit{V}}}^{(k)}) = \textbf{\textit{H}} \mathrm{diag}(\textbf{\textit{H}}^\top \overline{\textbf{\textit{V}}}^{(k)})$. Next, we stack each voltage coefficient matrix and current vector corresponding to a batch of measurements as follows: \begin{equation*}
     \textbf{\textit{A}} = \begin{bmatrix}
    \overline{\textbf{\textit{A}}}(\overline{\textbf{\textit{V}}}^{(1)})\\\overline{\textbf{\textit{A}}}(\overline{\textbf{\textit{V}}}^{(2)})\\\vdots\\\overline{\textbf{\textit{A}}}(\overline{\textbf{\textit{V}}}^{(T)})\end{bmatrix} \in \mathbb{C}^{nT \times e} \text{ and } \textbf{\textit{i}}=\begin{bmatrix} \overline{\textbf{\textit{I}}}^{(1)}\\
    \overline{\textbf{\textit{I}}}^{(2)}\\\vdots\\\overline{\textbf{\textit{I}}}^{(T)}\end{bmatrix} \in \mathbb{C}^{nT}.
     \end{equation*} 
Here, the $\textbf{\textit{A}}$ matrix is termed the \emph{average voltage coefficient matrix}, and $\textbf{\textit{i}}$ is the \emph{average current vector}. Then, we rewrite~\eqref{eq4} as
\begin{equation}\label{eq5} \textbf{\textit{i}}=\textbf{\textit{A}}\textbf{\textit{y}}. \end{equation}

\section{Problem Formulation}\label{Section3}

In this section, we consider graph rigidity theory and the plug-in estimation method to estimate an electric grid topology and admittance parameters using synchrophasor measurements.

\begin{problem}\label{p1}
   Consider an electric grid as defined by   $\mathcal{G}(\mathcal{N},\mathcal{E})$ with an unknown admittance vector $\textbf{\textit{y}} \in \mathbb{C}^{e}$. At each node $n\in \mathcal{N}$ and each operating condition $k\in\mathcal{T}$, we collect measurements of time-series current phasors $\textbf{\textit{I}}^{(k)}_1,\textbf{\textit{I}}^{(k)}_2,\dotsc,\textbf{\textit{I}}^{(k)}_B \in \mathbb{C}^{n}$ and the corresponding time-series voltage phasors $\textbf{\textit{V}}^{(k)}_1,\textbf{\textit{V}}^{(k)}_2,\dotsc,\textbf{\textit{V}}^{(k)}_B \in \mathbb{C}^{n}$. Suppose that $T$ operating condtions are considered, we seek to estimate the admittance vector $\textbf{\textit{y}}$ and the corresponding grid topology.
   \end{problem}

To obtain the solution to Problem~\ref{p1}, we leverage both graph rigidity theory and the plug-in estimation method. Graph rigidity theory provides a theoretical tool to determine the sufficiency of the measurements for the estimation, whereas the plug-in estimation method offers a computationally efficient and scalable solution to the problem.

\subsection{Graph Rigidity Theory}\label{section3a}

In our previous work \cite{11107735}, we investigated the problem of determining the minimum number of operating points (measurements) required for accurate estimation of both the topology and admittance parameters of an electric grid. To examine the conditions under which the parameters can be uniquely determined, we introduced concepts from graph rigidity theory, namely rigid frameworks and globally rigid frameworks \cite{Conditions, Globally, gortler2014generic, jordan2017global, jordan2016ii}.

We consider an electric grid which corresponds to a complete graph $\mathcal{G}(\mathcal{N,E})$. Since $\mathcal{G}(\mathcal{N,E})$ is a complete graph, $\mathcal{G}(\mathcal{N,E})$ is generically globally rigid \cite[Remark 5]{11107735}. The rigidity matrix of $\mathcal{G}(\mathcal{N,E})$ is represented by $\textbf{\textit{R}}\in \mathbb{C}^{e \times nT}$ \cite{11107735}. We established a connection between rigid frameworks and the admittance network modeled by the graph $\mathcal{G}(\mathcal{N}, \mathcal{E})$ via Lemma~\ref{l1}.

\begin{lemma}[{\cite[Lemma 3]{11107735}}]\label{l1}
     For the voltage coefficient matrix $\textbf{\textit{A}} \in \mathbb{C}^{nT \times e}$, where $T$ is the minimum number of required measurements, and $n$ is the number of network nodes, the rank of $\textbf{\textit{A}}$ is equal to the rank of $\textbf{\textit{R}}$ and is equal to $nT - T(T + 1)/2$ when $n\geq T$.
\end{lemma}

 Knowing the rank of $\textbf{\textit{A}}$, we derived Theorem~\ref{t1} that provides theoretical bounds on the number of measurements necessary for unique identification when the measurements have no noise. In the presence of noise, Theorem~\ref{t1} establishes the sufficiency of the measurements for accurate estimation. 

In what follows, we introduce a new vector and a new definition used in Theorem~\ref{t1}. A \emph{nodal voltage vector} for node $i\in\mathcal{N}$ is denoted by $\Tilde{\textbf{\textit{V}}}_i$ and defined by $\Tilde{\textbf{\textit{V}}}_i = [V_{i}^{(1)},V_{i}^{(2)},\dotsc, V_{i}^{(T)}]^\top \in \mathbb{C}^{\tau}$, where $\{1,2,\dotsc,T\} \in \mathcal{T}$. We also introduce the concept of genericity in order to find the minimum number of measurements $T$ such that the electric grid topology and admittance parameters can be uniquely determined. See~\cite{11107735} for further details on genericity. 

\begin{definition}{\cite[Definition 6]{gortler2014generic}}\label{d3}
     The nodal voltage vectors $[\Tilde{\textbf{\textit{V}}}_1, \Tilde{\textbf{\textit{V}}}_2,\dotsc, \Tilde{\textbf{\textit{V}}}_n]$ are \emph{generic} if they do not satisfy any non-zero algebraic equation with rational coefficients. 
\end{definition} 

\begin{remark}
If the nodal voltage vectors $[\Tilde{\textbf{\textit{V}}}_1, \Tilde{\textbf{\textit{V}}}_2,\dotsc, \Tilde{\textbf{\textit{V}}}_n]$ are linearly dependent, then they are not generic.
\end{remark}

 \begin{theorem}[{\cite[Section III-B]{11107735}}]\label{t1}
 For an electric grid represented by a complete graph $\mathcal{G}(\mathcal{N},\mathcal{E})$,
 assume the nodal voltage vectors $[\Tilde{\textbf{\textit{V}}}_1, \Tilde{\textbf{\textit{V}}}_2,\dotsc, \Tilde{\textbf{\textit{V}}}_n]$ are generic. 
 Then, the minimum number of phasor-based measurements $T$ such that $\textbf{\textit{i}}=\mathcal{\textbf{A}}\textbf{\textit{y}}$ has a unique solution $\textbf{\textit{y}}$ is $T = n-1$.
\end{theorem}

Theorem~\ref{t1} characterizes measurement sufficiency for accurate topology and admittance estimation. In our prior work~\cite{ElectricGrid}, we considered a structured total least squares problem to estimate the topology and admittance parameters in the presence of noisy synchrophasor measurements. The measurement noises in the current and voltage vectors at each operating condition $k\in\mathcal{T}$ are denoted by $\Delta \textbf{\textit{I}}^{(k)} \in \mathbb{C}^n$ and $\Delta \textbf{\textit{V}}^{(k)} \in \mathbb{C}^n$, respectively. We define a noise vector $\textbf{\textit{s}}^{(k)} := [(\Delta \textbf{\textit{V}}_{Re}^{(k)})^\top, (\Delta \textbf{\textit{V}}_{Im}^{(k)})^\top, (\Delta \textbf{\textit{I}}_{Re}^{(k)})^\top, (\Delta \textbf{\textit{I}}_{Im}^{(k)})^\top]^\top, \forall k \in \mathcal{T}$, where the subscripts $Re$ and $Im$ represent the real and imaginary parts of the corresponding complex matrices, respectively. If the $\textbf{\textbf{s}}^{(k)}$ are regarded as deterministic unknowns, the structured total least squares problem can be formulated as an equality constrained optimization problem as follows:
\begin{multline}\label{eq6}
\min_{(\textbf{\textit{s}}^{(k)})_{k=1}^{\tau},\textbf{\textit{y}}_{Re},\textbf{\textit{y}}_{Im}} \frac{1}{2} \sum_{k=1}^{\tau} (\textbf{\textit{s}}^{(k)})^\top W \textbf{\textit{s}}^{(k)} \\ \footnotesize
\textrm{s.t.}
\footnotesize
    \begin{bmatrix}
        \textbf{\textit{I}}_{Re}^{(k)} \\
        \textbf{\textit{I}}_{Im}^{(k)}
    \end{bmatrix} + \begin{bmatrix}
        \Delta\textbf{\textit{I}}_{Re}^{(k)} \\
        \Delta\textbf{\textit{I}}_{Im}^{(k)}
    \end{bmatrix} = \left(
    \begin{bmatrix}
        \textbf{\textit{H}} \mathrm{diag}(\textbf{\textit{H}}^\top\textbf{\textit{V}}_{Re}^{(k)}) & -\textbf{\textit{H}} \mathrm{diag}(\textbf{\textit{H}}^\top\textbf{\textit{V}}_{Im}^{(k)}) \\ \textbf{\textit{H}} \mathrm{diag}(\textbf{\textit{H}}^\top\textbf{\textit{V}}_{Im}^{(k)}) & \textbf{\textit{H}} \mathrm{diag}(\textbf{\textit{H}}^\top\textbf{\textit{V}}_{Re}^{(k)})
    \end{bmatrix}\right. \\ \footnotesize + \left.\begin{bmatrix}     \textbf{\textit{H}} \mathrm{diag}(\textbf{\textit{H}}^\top\Delta\textbf{\textit{V}}_{Re}^{(k)}) & -\textbf{\textit{H}} \mathrm{diag}(\textbf{\textit{H}}^\top\Delta\textbf{\textit{V}}_{Im}^{(k)}) \\ \textbf{\textit{H}} \mathrm{diag}(\textbf{\textit{H}}^\top\Delta\textbf{\textit{V}}_{Im}^{(k)}) & \textbf{\textit{H}} \mathrm{diag}(\textbf{\textit{H}}^\top\Delta\textbf{\textit{V}}_{Re}^{(k)}) \end{bmatrix} \right) \begin{bmatrix} \textbf{\textit{y}}_{Re} \\ \textbf{\textit{y}}_{Im} \end{bmatrix}, \forall k \in \mathcal{T},\end{multline}
where $W$ is a weighting matrix in $\mathbb{R}^{2(n+e)\times 2(n+e)}$ that is designed to penalize the error terms.
 
Nonetheless, the structured total least squares problem~\eqref{eq6}
cannot be readily solved using off-the-shelf solvers and standard computing hardware for networks exceeding $34$ nodes, corresponding to $561$ unknown admittance parameters. This scalability issue arises primarily from the increasing computational complexity and memory demands associated with solving large-scale, high-dimensional optimization problems involving large matrices. In practical settings, electric networks often contain 100 nodes or more, making the application of the original method computationally intractable. To address this challenge, we propose integrating the plug-in estimation method~\cite{wright2011using} into our existing approach, which offers improved scalability of the problem.


\subsection{Plug-in Estimation Method}\label{section3b}

In the context of topology and admittance estimation, the plug-in estimation method~\cite{wright2011using} determines grid topology and admittance parameters by leveraging averaged time-series voltage and current phasor measurements. We start by obtaining intermediate quantities such as the voltage coefficient matrix $\overline{\textbf{\textit{A}}}(\overline{\textbf{\textit{V}}}^{(k)})$ and current vector $\overline{\textbf{\textit{I}}}^{(k)}$. We obtain $\overline{\textbf{\textit{A}}}(\overline{\textbf{\textit{V}}}^{(k)})$ and $\overline{\textbf{\textit{I}}}^{(k)}$ by averaging multiple time-series voltage measurements $\textbf{\textit{V}}^{(k)}_1,\textbf{\textit{V}}^{(k)}_2,\dotsc,\textbf{\textit{V}}^{(k)}_B$ and current measurements $\textbf{\textit{I}}^{(k)}_1,\textbf{\textit{I}}^{(k)}_2,\dotsc,\textbf{\textit{I}}^{(k)}_B$ at the same operating condition $k\in\mathcal{T}$. Once obtained, the averaged quantities $\overline{\textbf{\textit{A}}}(\overline{\textbf{\textit{V}}}^{(k)})$ and $\overline{\textbf{\textit{I}}}^{(k)}$ are concatenated for all $k\in \mathcal{T}$ to obtain $\textbf{\textit{A}}$ and $\textbf{\textit{i}}$, and then ``plugged in" to the estimation algorithm to estimate the admittance parameters and infer the topology. As established in Theorem~\ref{t1}, we require data from at least $n-1$ distinct operating conditions to estimate the admittance parameters.

In practice, gathering the $n-1$ independent operating conditions (required by Theorem~\ref{t1}) purely through natural load fluctuations is difficult. Since grid voltages are tightly regulated, the time-series voltage measurements are dominated by the nominal voltage profile, making the voltage vectors highly collinear with $\mathbf{1}_n$. Consequently, the subtle voltage variations driven by load changes are obscured by this static component, yielding ill-conditioned voltage coefficient matrix $\overline{\textbf{\textit{A}}}(\overline{\textbf{\textit{V}}}^{(k)})$ that numerical solvers often treat as rank-deficient. To mitigate this numerical ill-conditioning and expose the true linear independence of the operating conditions, we preprocess the voltage measurements using a voltage-centering transformation. That is, the transformation relies on the zero-row-sum property of the admittance matrix, $\textbf{\textit{Y}}\mathbf{1}_n = \textbf{0}_n$. Due to this property, current injections remain invariant under a uniform voltage shift, making the fundamental relationship $\textbf{\textit{I}} = \textbf{\textit{Y}}\textbf{\textit{V}}$ mathematically equivalent to $\textbf{\textit{I}} = \textbf{\textit{Y}}(\textbf{\textit{V}} - V_{\text{avg}}\mathbf{1}_n)$, where $V_{\text{avg}} = \frac{\mathbf{1}_n^\top \textbf{\textit{V}}}{n}$. Leveraging this property, for every operating condition $k \in \mathcal{T}$ and every batch time-series voltage measurement $t\in\{1,2,\dotsc,B\}$, we substitute the voltage vector $\textbf{\textit{V}}_{t}^{(k)}$ with a \emph{centered voltage vector} denoted by 
\begin{equation}\label{equation8}{\textbf{\textit{V}}'}_{t}^{(k)} = \textbf{\textit{V}}_{t}^{(k)} - V_{\text{avg},t}^{(k)}\mathbf{1}_n,
\end{equation} where $V_{\text{avg},t}^{(k)}$ is the scalar average of the voltages in $\textbf{\textit{V}}_{t}^{(k)}$. Accordingly, at operating condition $k\in\mathcal{T}$, the voltage coefficient matrix  $\overline{\textbf{\textit{A}}}(\overline{\textbf{\textit{V}}}^{(k)})$ becomes $\overline{\textbf{\textit{A}}'}(\overline{\textbf{\textit{V}}'}^{(k)}) = \textbf{\textit{H}} \mathrm{diag}(\textbf{\textit{H}}^\top \overline{\textbf{\textit{V}}'}^{(k)})$. Although the voltage-centering transformation does not change the branch voltage drops—meaning the coefficient matrix $\overline{\textbf{\textit{A}}'}(\overline{\textbf{\textit{V}}'}^{(k)})$ is theoretically identical to $\overline{\textbf{\textit{A}}}(\overline{\textbf{\textit{V}}}^{(k)})$—it fundamentally improves the numerical conditioning of the problem. By removing the static nominal voltage profile, the transformation isolates the informative, load-driven fluctuations. This significantly reduces the condition number of the voltage coefficient matrices, ensuring the estimation remains solvable even when natural load variations are minimal.

The final stage of the method involves estimating the admittance vector via least-squares regression and isolating the true network edges through thresholding. Using the average voltage coefficient matrix $\textbf{\textit{A}}$ and the average current vector $\textbf{\textit{i}}$, we determine the \emph{estimated admittance vector}, $\hat{\textbf{\textit{y}}}= [\hat{y}_{1,2}, \hat{y}_{1,3}, \dotsc, \hat{y}_{n-1,n}]$, by solving a least-squares regression problem as follows:
\begin{equation}\label{eq7}
 \hat{\textbf{\textit{y}}} = \arg\min_{\textbf{\textit{y}}} \|\textbf{\textit{A}}\textbf{\textit{y}} - \textbf{\textit{i}}\|_2^2.
\end{equation} 
Next, we infer the \emph{estimated grid topology}, $\mathcal{\hat{G}}(\mathcal{N},\hat{\mathcal{E}})$. Since finding the topology primarily means identifying the active set of edges $\hat{\mathcal{E}} \subseteq \mathcal{E}$, we apply a thresholding operation to $\hat{\textbf{\textit{y}}}$. By setting any entries of $\hat{\textbf{\textit{y}}}$ below a predefined threshold $\epsilon$ to zero, we remove the inactive connections in the network. This step yields a sparse admittance vector $\hat{\textbf{\textit{y}}} \in \mathbb{C}^e$ and its corresponding estimated edge set $\hat{\mathcal{E}}$.

The plug-in estimation method allows the averaging step and the least squares regression step to be computed separately. Furthermore, it avoids direct inversion of large-scale matrices or solving high-dimensional optimization problems that often hinder scalability. Consequently, the method provides a computationally efficient and scalable alternative to the structure total least squares method \cite{ElectricGrid}.

In Algorithm~\ref{alg1}, we provide an overview of the plug-in estimation method, where we determine the 
\emph{estimated grid topology}, $\mathcal{\hat{G}} (\mathcal{{N},\hat{E}})$ and the \emph{estimated admittance vector}, $\hat{\textbf{\textit{y}}}$, using noisy measurements. 
Without prior topology information, we represent the grid topology to be estimated by a complete graph $\mathcal{G}$. In Algorithm~\ref{alg1}~(Line~\ref{line1}), we construct the incidence matrix $\textbf{\textit{H}}$ corresponding to the complete graph, $\mathcal{G}$, as defined in~\eqref{equation1}. In Line~\ref{line2}-\ref{line10}, we compute the voltage coefficient matrix $\overline{\textbf{\textit{A}}'}(\overline{\textbf{\textit{V}}'}^{(k)})$ and current vector $\overline{\textbf{\textit{I}}}^{(k)}$ for all $k\in\{1,2,\dotsc,T\}$. More specifically, in Lines~\ref{line5}, for each time-series measurement $t\in\{1,2,\dotsc,B\}$, we substitute the voltage vector $\textbf{\textit{V}}_{t}^{(k)}$ with a centered voltage vector ${\textbf{\textit{V}}'}_{t}^{(k)}$ and we compute the average centered voltage vector $\overline{\textbf{\textit{V}}'}^{(k)}$ and average current vector $\overline{\textbf{\textit{I}}}^{(k)}$ over all time-series measurements $t\in\{1,2,\dotsc,B\}$ as in Line~\ref{line9}. In Line~\ref{line11} and Line~\ref{line12}, we concatenate the voltage coefficient matrix $\overline{\textbf{\textit{A}}'}(\overline{\textbf{\textit{V}}'}^{(k)})$ and current vector $\overline{\textbf{\textit{I}}}^{(k)}$ for all $k\in\mathcal{T}$ to obtain $\textbf{\textit{A}}$ and $\textbf{\textit{i}}$, respectively. In Line~\ref{line13}, we solve the least-squares regression problem~\eqref{eq7} to obtain $\hat{\textbf{\textit{y}}}$. In Lines~\ref{line14}-\ref{line16} we set to zero the admittance parameters with a magnitude that is less than a design threshold, $\epsilon$. In Line~\ref{line17}, we return the estimated admittance vector $\hat{\textbf{\textit{y}}}$. In Line~\ref{line18}, we determine the set of edges with nonzero admittance parameters $\hat{\mathcal{E}}$. In Line~\ref{line19}, we return the estimated topology $\hat{\mathcal{G}}(\mathcal{N},\hat{\mathcal{E}})$.

\begin{algorithm}[ht]
\caption{Plug-in Estimation of Admittance Parameters and Corresponding Topology of an Electric Grid.}
\label{alg1}
\begin{algorithmic}[1]
\Require $n$, $B$, $T$, $\epsilon = 4$, $\{\{\textbf{\textit{V}}^{(k)}_t,\textbf{\textit{I}}^{(k)}_t \in \mathbb{C}^n\}_{t=1}^B\}_{k=1}^T$. 
\State Construct $\textbf{\textit{H}} \in \mathbb{R}^{n\times e}$.\label{line1}
\For{$k = 1$ to $T$}\label{line2}
\State $\overline{\textbf{\textit{V}}'}^{(k)} \gets \mathbf{0}_n$, $\overline{\textbf{\textit{I}}}^{(k)} \gets \mathbf{0}_n$,\label{line3}
\For{$t = 1$ to $B$}\label{line4}
    \State ${\textbf{\textit{V}}'}_{t}^{(k)} = \textbf{\textit{V}}_{t}^{(k)} - V_{\text{avg},t}^{(k)}\mathbf{1}_n$,\label{line5}
    \State $\overline{\textbf{\textit{V}}'}^{(k)} \gets \overline{\textbf{\textit{V}}'}^{(k)} + \textbf{\textit{V}}'^{(k)}_t$,\label{line6}
    \State $\overline{\textbf{\textit{I}}}^{(k)} \gets \overline{\textbf{\textit{I}}}^{(k)} + \textbf{\textit{I}}^{(k)}_t$,\label{line7}
\EndFor \label{line8}
\State $\overline{\textbf{\textit{V}}'}^{(k)} \gets \frac{1}{B}\overline{\textbf{\textit{V}}'}^{(k)}$, $\overline{\textbf{\textit{I}}}^{(k)} \gets \frac{1}{B}\overline{\textbf{\textit{I}}}^{(k)}$, \newline \hspace*{1.2em} $\overline{\textbf{\textit{A}}'}(\overline{\textbf{\textit{V}}'}^{(k)}) \gets \textbf{\textit{H}} \mathrm{diag}(\textbf{\textit{H}}^\top \overline{\textbf{\textit{V}}'}^{(k)})$.\label{line9}
\EndFor \label{line10}
\State $\textbf{\textit{A}} = \left[{\overline{\textbf{\textit{A}}'}(\overline{\textbf{\textit{V}}'}^{(1)})}^\top, {\overline{\textbf{\textit{A}}'}(\overline{\textbf{\textit{V}}'}^{(2)})}^\top, \dotsc, {\overline{\textbf{\textit{A}}'}(\overline{\textbf{\textit{V}}'}^{(T)})}^\top\right]^\top$.\label{line11}
\State $\textbf{\textit{i}}=\left[{\overline{\textbf{\textit{I}}}^{(1)}}^\top, {\overline{\textbf{\textit{I}}}^{(2)}}^\top, \dotsc, {\overline{\textbf{\textit{I}}}^{(T)}}^\top\right]^\top$.\label{line12}
\State Solve equation~\eqref{eq7}: $\hat{\textbf{\textit{y}}} \gets \arg\min_{\textbf{\textit{y}}} \|\textbf{\textit{A}}\textbf{\textit{y}} - \textbf{\textit{i}}\|_2^2$.\label{line13}
\For{edge $(i,j) = 1:e$}\label{line14}
    \State $|\hat y_{i,j}|<\epsilon \;\rightarrow\; \hat y_{i,j}=0$.\label{line15}
\EndFor\label{line16}
\State \Return $\hat{\textbf{\textit{y}}}= [\hat{y}_{1,2}, \hat{y}_{1,3}, \dotsc, \hat{y}_{n-1,n}]$.\label{line17}
\State $\hat{\mathcal{E}} \gets \{ (i,j): \hat{y}_{i,j} \neq 0 \}$.\label{line18}
\State \Return $\hat{\mathcal{G}}(\mathcal{N},\hat{\mathcal{E}})$.\label{line19}  
\end{algorithmic}
\end{algorithm}

In Algorithm~\ref{alg1}, we choose the threshold $\epsilon$ to distinguish true physical line connections from noise-induced errors. If $\epsilon$ is too large, we risk over-pruning lines with low admittance. Similarly, if $\epsilon$ is too small we risk misclassifying grid edges due to non-zero error residuals. Accordingly, the threshold $\epsilon$ must be strictly bounded such that $\sigma_{\text{noise}} < \epsilon < |y_{\text{min}}|$, where $|y_{\text{min}}|$ is the minimum absolute admittance magnitude in the physical grid, and $\sigma_{\text{noise}}$ is the estimation error variance. In practice, we envision that $\epsilon$ will be chosen using historical PMU measurements to ensure it exceeds the measurement error level while remaining below the lowest line admittance parameter in the grid.

\section{Numerical Simulations}\label{section4}
We evaluate the performance of the proposed plug-in estimation method via extensive numerical simulations on standard IEEE test feeders~\cite{IEEE}. Our goal is to assess the method's accuracy, computational cost, robustness to noise, scalability, and performance with partial topology knowledge. We consider the following IEEE test feeders.

\begin{itemize}
\item \textbf{IEEE 13-node feeder}, modeled as a 12-node balanced single-phase system (ignore the substation node). We use this feeder to demonstrate the performance of the plug-in estimation method in comparison to the structured total least squares (STLS) method.
\item \textbf{IEEE 123-node feeder}, modeled as a 122-node balanced single-phase system (ignore the substation node). We simulate this feeder to assess the scalability of the proposed method while ensuring that there is no prior topology information. 
\item \textbf{IEEE European low voltage test feeder}, a large low-voltage distribution feeder consists of 907 nodes which is representative of a real-world power system. We select this feeder to evaluate the scalability of the proposed method on an even larger network with the help of some prior topology information.
\end{itemize}

\subsection{Comparison with STLS}\label{section4a}
To benchmark our method, we compare its performance with the structured total least squares method proposed in our prior work \cite{ElectricGrid} on a modified version of the IEEE 13-node feeder \cite{IEEE}. 
We modify the IEEE 13-node feeder by removing the substation node, replacing the transformer between nodes 5 and 6, and the switch between nodes 7 and 10 with the admittance reported between nodes 2 and 5, and between nodes 10 and 11, respectively. We also remove the capacitors between nodes 9 and 11 as done in \cite{7741261}.

In what follows, the $12$ single phase nodes in the modified IEEE 13-node feeder are represented by a complete graph. That is, we consider an admittance network defined by $\mathcal{G}(\mathcal{N},\mathcal{E})$, with $n = 12$. To estimate the topology and admittance parameters of the modified IEEE 13-node feeder, we first collect $T$ voltage and current phasor measurements at each of the $12$ nodes.
 
To construct the voltage and current measurements at each of the $12$ node, our approach is as follows. The first voltage vector corresponding to $k = 1$, $\textbf{\textit{V}}^{(1)}$, is as reported in~\cite{IEEE}. For all subsequent operating conditions operating conditions $(k>1)$, we simulate the voltage vectors $\textbf{\textit{V}}^{(k)}$ using the Open Distribution System Simulator (OpenDSS)~\cite{EPRI2022}, driven by load profiles from the NextGen dataset~\cite{shaw2019nextgen}. In practice, the voltage measurements $\textbf{\textit{V}}^{(1)},\textbf{\textit{V}}^{(2)},\dotsc,\textbf{\textit{V}}^{(T)}$ are associated with instrument transformers~\cite{brady2020phasor}, which introduce minor reading errors despite being designed to protect voltage sensors. To accurately model these errors across all operating conditions, we add Gaussian noise to our measurements, specifically using a normal distribution with zero mean and a variance of $0.0001$. Next, we introduce measurement noise to the time-series voltage at $k\in\mathcal{T}$ denoted by $\textbf{\textit{V}}^{(k)}_t$, where $t = \{1,2,\dotsc,B\}$. Specifically, we represent the measurement noise by a Gaussian noise with zero mean and a signal to noise ratio (SNR) of 45 dB which is roughly translated to a variance of $\sigma = 0.005$~\cite{7741972}. That is, the time-series voltage vector at each operating condition $k\in \mathcal{T}$, $\textbf{\textit{V}}^{(k)}_t$, are obtained by adding the voltage vector at operating condition $k$, $\textbf{\textit{V}}^{(k)}$, to a vector of pseudorandom values between $-0.005|\textbf{\textit{V}}^{(k)}|$ and $0.005|\textbf{\textit{V}}^{(k)}|$. For example, at the first operating condition $k=1$, $\textbf{\textit{V}}^{(1)}_t \approx \textbf{\textit{V}}^{(1)} \pm 0.005|\textbf{\textit{V}}^{(1)}|$ for all $t \in \{1,2,\dotsc,B\}$. We then substitute the time-series voltage vector $\textbf{\textit{V}}_{t}^{(k)}$ with the centered voltage vector ${\textbf{\textit{V}}'}_{t}^{(k)}$ as in~\eqref{equation8} for all $k\in\mathcal{T}$ and for all $t\in \{1,2,\dotsc,B\}$.  The current vectors $\textbf{\textit{I}}^{(k)}$ are obtained by solving equation~\eqref{eq1} with $\textbf{\textit{Y}}$ being the actual admittance matrix as reported in~\cite{IEEE} and $\textbf{\textit{V}}^{(k)}$ as constructed above. Accordingly, the measurement noise is also applied to the time-series current vectors $\textbf{\textit{I}}^{(k)}_t$.

We then proceed to estimate the topology and admittance parameters of the feeder. After computing the averages and concatenating the matrices and vectors as in Lines~\ref{line11}-\ref{line12} of Algorithm~\ref{alg1}, we use MATLAB to solve the least-squares regression as in Line~\ref{line13} of Algorithm~\ref{alg1}. A threshold of $\epsilon = 4$ is used to remove negligible admittance values and extract the estimated admittance vector $\hat{\textbf{\textit{y}}}$, and the corresponding topology $\hat{\mathcal{G}}(\mathcal{N},\hat{\mathcal{E}})$. The simulations were conducted on a MacBook Pro with 16GB RAM and 2.3 GHz Quad-Core Intel Core i7 processor.

\begin{figure}[!t]
\centering
\includegraphics[width=0.495\textwidth]{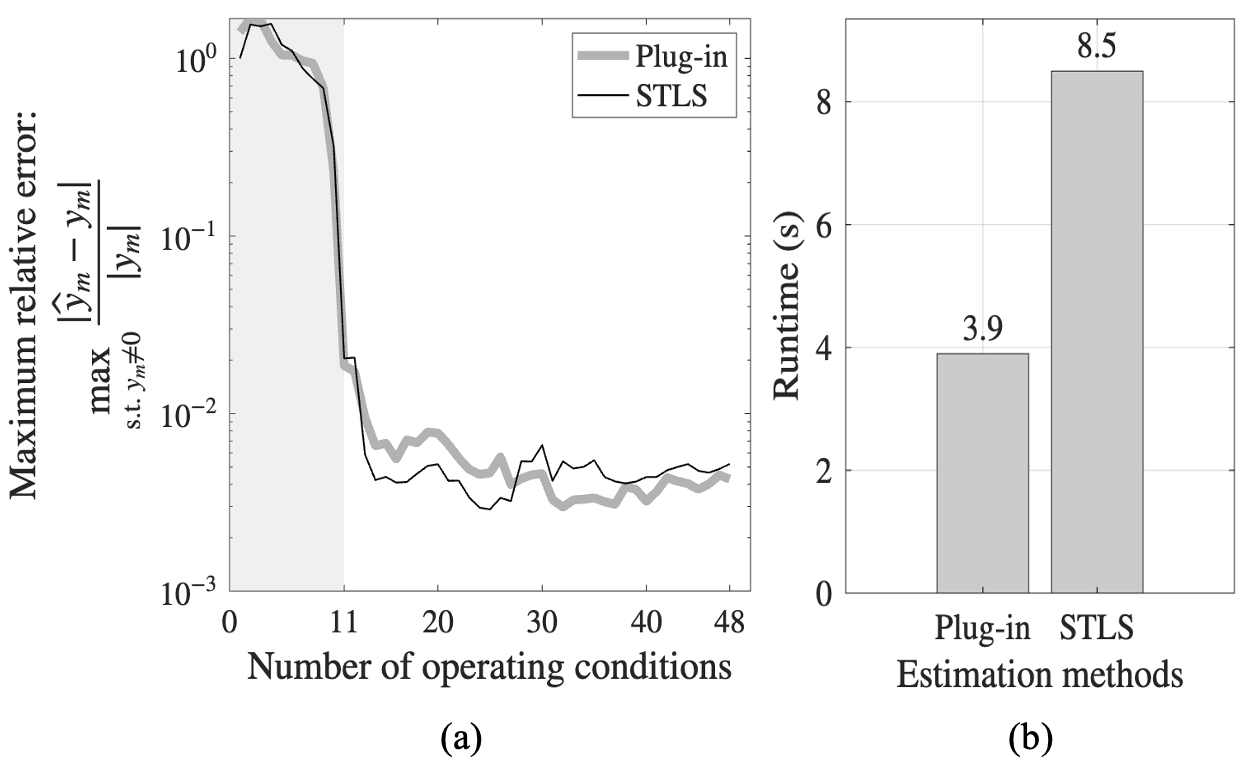}
\caption{Comparison between the STLS method and the plug-in estimation method on the IEEE 13-node feeder in terms of: (a) the accuracy; and (b) the total computational time. Shaded area in (a) indicates insufficient number of operating conditions leading to high estimation error.}
\label{Fig2}
\end{figure}

In Fig.~\ref{Fig2}, we present the comparison between the STLS method and the plug-in estimation method on the IEEE 13-node feeder in terms of: (a) the accuracy and (b) the computational time. The batch size of the time-series measurements considered in Fig.~\ref{Fig2} is $B = 1000$. According to Fig.~\ref{Fig2} (a), we observe a large maximum relative error in the admittance $y$ of both the STLS and the plug-in method when the number of operating conditions is less than the minimum requirement as per Theorem~\ref{t1}. Additionally, the error decreases with an increased number of operating condition. When the number of operating conditions is above the minimum requirement, the maximum relative error in the STLS method is similar to that of the plug-in method. In contrast, according to Fig.~\ref{Fig2} (b), we observe that the STLS incurs a computational cost approximately 2 times greater than that of the plug-in method. This comparison highlights the computational advantages of the plug-in method over the STLS method.

\begin{figure}[!t]
\centering
\includegraphics[width=0.47\textwidth]{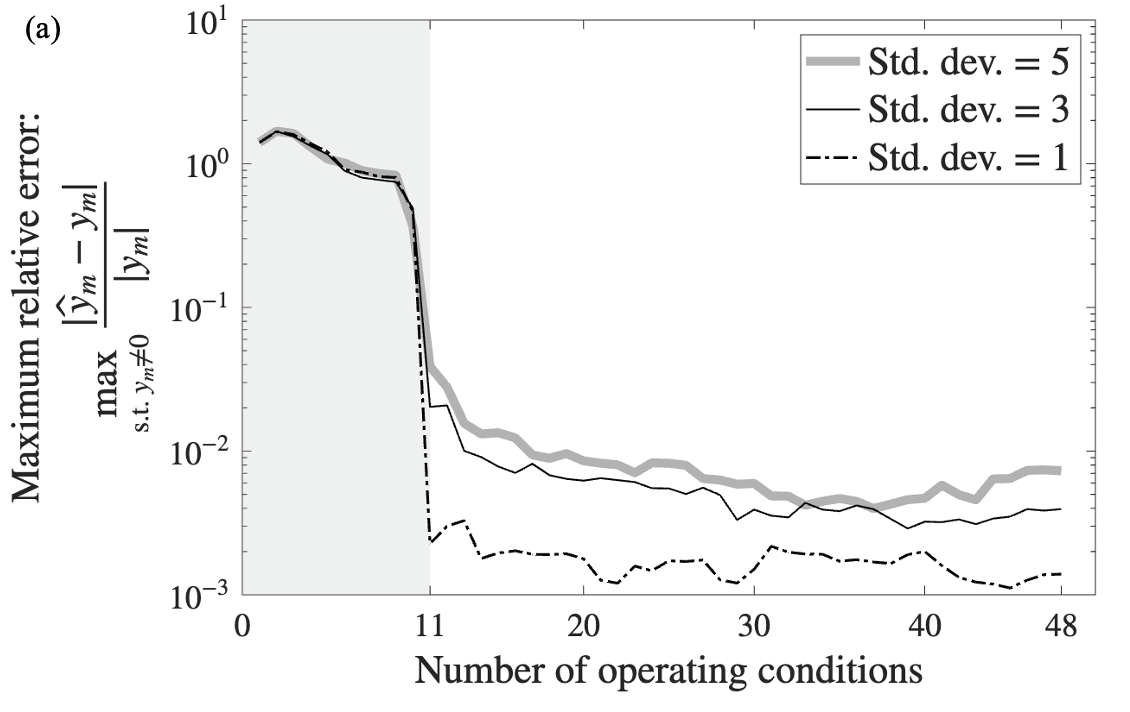}
\includegraphics[width=0.47\textwidth]{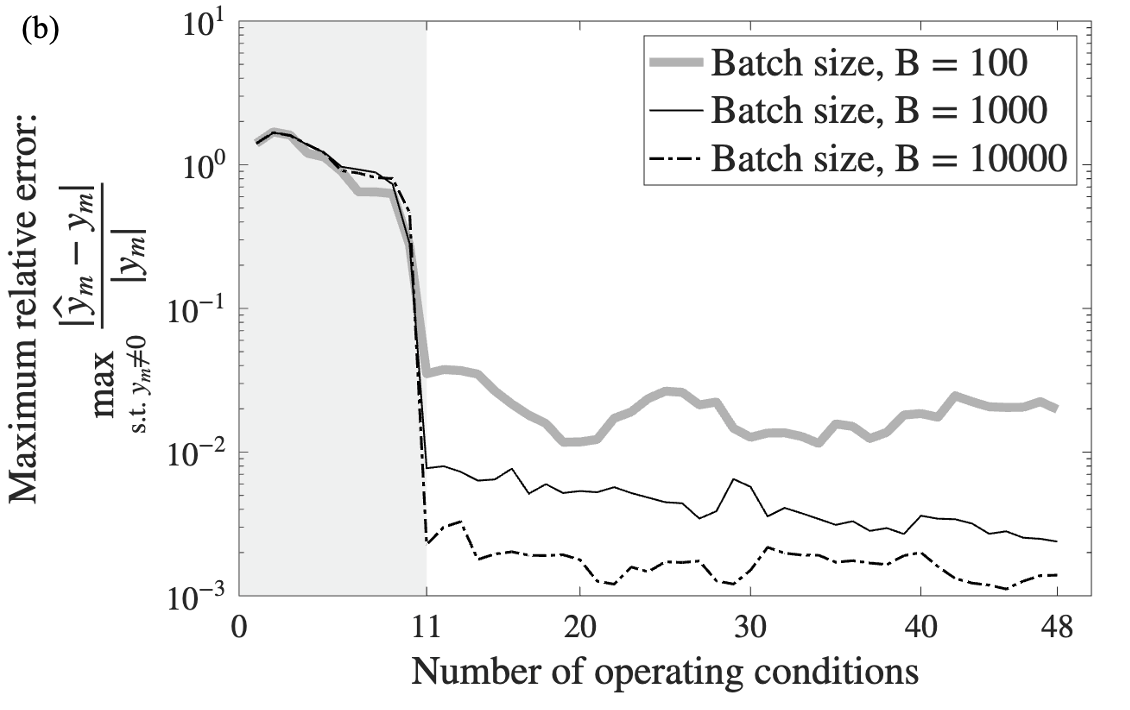}
\caption{Maximum relative error between the estimated and the actual admittance parameters of the IEEE 13-node feeder across: (a) different noise levels with the batch size, $B = 10000$; (b) different batch sizes at the noise level with the standard deviation = 1. Shaded area indicates insufficient
number of operating conditions leading to high estimation error.}
\label{Fig3}
\end{figure}

In Fig.~\ref{Fig3}, we conduct simulations of the plug-in estimation method on the IEEE-13 node feeder across various noise levels and different sizes of the batch time-series measurements. The noise levels in Fig.~\ref{Fig3}~(a) are modeled as zero-mean Gaussian with varying standard deviations of $1$, $3$, and $5$. According to Fig.~\ref{Fig3}~(a), the error increases with higher noise levels as expected. Furthermore, to examine the impact of batch sizes on the estimation, the batch sizes considered in Fig.~\ref{Fig3}~(b) are selected from a set of $\{100, 1000, 10000\}$. According to Fig.~\ref{Fig3}~(b), the plug-in estimation method performs better with an increased batch size.

\subsection{Scalability}\label{section4b}

To assess the scalability of the proposed method, we applied the plug-in estimation method to the IEEE 123-node feeder~\cite{IEEE}. This feeder presents a significantly larger and more complex topology compared to the IEEE 13-node feeder, providing a suitable benchmark to evaluate computational performance and estimation accuracy at a larger scale. Simulations were conducted without any prior topology information, with phasor measurements data obtained in a similar pattern as in Section~\ref{section4a}.

\begin{figure}[!t]
\centering
\includegraphics[width=0.48\textwidth]{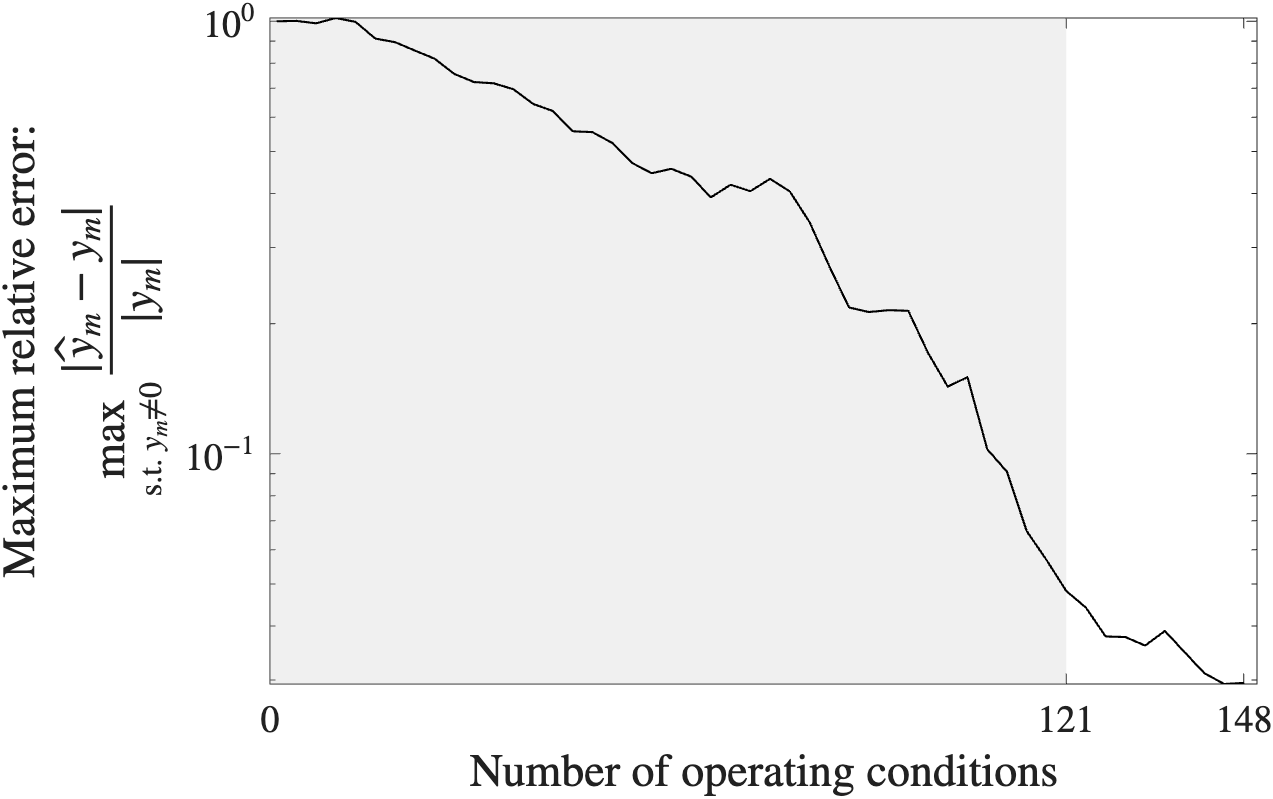}
\caption{Maximum relative error between the estimated and actual admittance parameters of the IEEE 123-node feeder with the batch size, $B = 1000$. Shaded area indicates insufficient number of operating conditions leading to high estimation error.}
\label{Fig4}
\end{figure}

Figure~\ref{Fig4} presents the maximum relative error between the estimated and actual admittance parameters of the IEEE 123-node feeder. The simulation was done on a batch size, $B = 1000$. According to Fig~\ref{Fig4}, the result indicates that the plug-in estimation method maintains high accuracy even as network size increases, provided that a sufficient number of operating conditions is available. The result in Fig.~\ref{Fig4} shows the applicability of the method on electric grids comprising hundreds of nodes without prior topology information.

Table~\ref{tab1} presents the number of misidentified edges after running the estimation algorithm on the IEEE 13-node and IEEE 123-node feeders. In Table~\ref{tab1}, we analyze the impact of different sampling characteristics, including different batch sizes and different standard deviations, on the topology estimation accuracy. According to Table~\ref{tab1}, the algorithm performs accurately across all scenarios, with no misidentification for the IEEE 13-node. However, for the IEEE 123-node, when the sample size $B = 100$, the algorithm misidentifies 18 and 326 edges for the noise levels with standard deviation of 3 and 5, respectively. Note that, in Algorithm~\ref{alg1}, we consider 48 operating conditions for the IEEE 13-node and 148 for the 123-node. We envision that having additional operating conditions will decrease the number of misidentified edges.

\begin{table}[ht]
\caption{Number of misidentified edges after running the estimation algorithm on the IEEE 13-node and IEEE 123-node.}
    \label{tab1}
\centering
\begin{tabular}{|c|c|c|c|}
\hline
\multicolumn{2}{|c|}{Sampling} & \multicolumn{2}{|c|}{Number of misidentified edges} \\ \cline{3-4} 
\multicolumn{2}{|c|}{characteristics} & IEEE 13-node & IEEE 123-node \\
\hline
\multirow{3}{6em}{Standard deviation = 1} & B = 100 & 0/66 & 0/7381 \\ \cline{2-4}
& B = 1000 & 0/66 & 0/7381 \\ \cline{2-4}
& B = 10000 & 0/66 & 0/7381 \\ 
\hline
\multirow{3}{6em}{Standard deviation = 3} & B = 100 & 0/66 & 18/7381 \\ \cline{2-4}
& B = 1000 & 0/66 & 0/7381 \\ \cline{2-4}
& B = 10000 & 0/66 & 0/7381 \\ 
\hline
\multirow{3}{6em}{Standard deviation = 5} & B = 100 & 0/66 & 326/7381 \\ \cline{2-4}
& B = 1000 & 0/66 & 0/7381 \\ \cline{2-4}
& B = 10000 & 0/66 & 0/7381 \\ 
\hline
\end{tabular}
\end{table}

\subsection{Admittance estimation with prior topology information}\label{section4c}

In practical settings, the assumption that there is no prior topology information—and that the network must be represented by a complete graph—is often unnecessary. Instead, partial knowledge of the grid topology, such as a subset of known connections from utility records, can be leveraged to enhance the estimation method. To demonstrate this scenario, we adapt the plug-in estimation method to incorporate known edges information by fixing specific entries in the admittance vector and applying our estimation method and threshold filter exclusively to the unknown elements.

We validate this scenario on the IEEE European low voltage test feeder~\cite{IEEE}. This feeder presents an even larger distribution system consisting of $907$ nodes, making it highly representative of a real-world power grid. To model the availability of prior topology information, we start with the true radial topology of $906$ edges and introduce $500$ ``false candidate" edges to represent the unknown connections. As the feeder comprises $907$ nodes, a complete graph lacking prior topology data would require estimating $\frac{907 \times 906}{2} = 410 871$ potential edges. By leveraging prior topology information and limiting the uncertainty to $500$ candidate edges, the total number of edges requiring estimation is drastically reduced from $410 871$ to $1406$. This reduction not only reflects realistic scenarios where utilities possess partial network data but also significantly mitigates computational complexity and memory constraints. The topological data for the IEEE European Low Voltage Test Feeder is obtained from~\cite{IEEE}, whereas the phasor measurements data is constructed in a similar pattern as in Section~\ref{section4a}.

\begin{figure}[!t]
\centering
\includegraphics[width=0.48\textwidth]{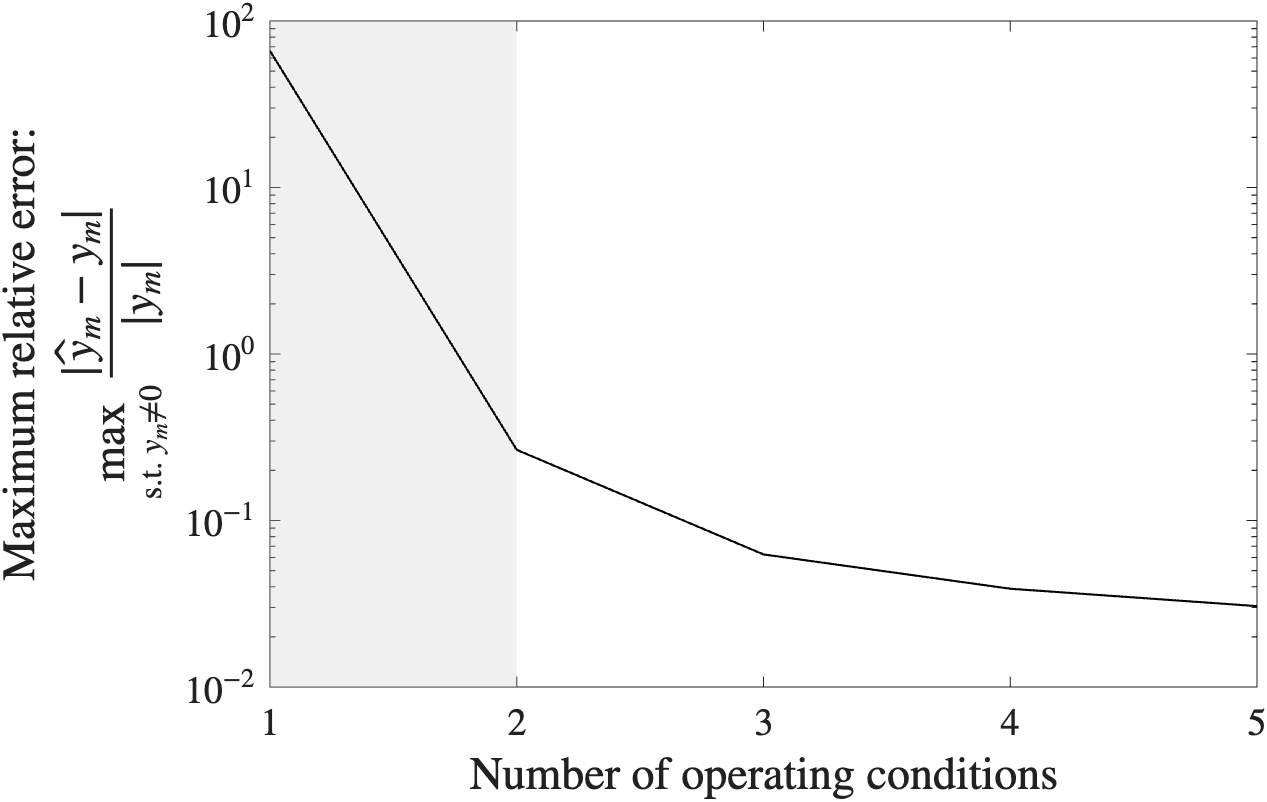}
\caption{Maximum relative error between the estimated and actual admittance parameters of the IEEE European low
voltage test feeder with the batch size, $B=1000$.}
\label{Fig5}
\end{figure}

In Fig~\ref{Fig5}, we present the maximum relative error between the estimated and actual admittance parameters of the IEEE European low voltage test feeder. According to Fig~\ref{Fig5}, the error decreases in accordance with the number of operating conditions as in the previous cases. Note that, since the number of edges requiring estimation drops to $1406$, only two operating conditions is sufficient for an accurate estimation. The result in Fig.~\ref{Fig5} demonstrates the method's applicability to power systems possessing partial knowledge of the grid topology, enabling practical deployment in utility networks even larger than those evaluated in Section~\ref{section4a} and~\ref{section4b}.

\section{Conclusion}\label{section5}

We presented a scalable approach to topology and admittance estimation in power systems using a plug-in estimation method. The plug-in estimation method significantly reduces computational time by separating the estimation problem into two stages: (i) constructing and averaging voltage coefficient matrices and current vectors from phasor data, and (ii) estimating admittance parameters via a least-squares regression with thresholding to recover the network topology. Consequently, the method avoids large matrix inversion and iterative optimization, as is otherwise required by a structured total least squares method. By means of numerical simulations on the IEEE 13-node, IEEE 123-node, and IEEE European low voltage test feeders, we observe our plug-in estimation method outperforms a structured total least squares method in terms of scalability to larger power grids and computational time.

\bibliographystyle{IEEEtran}
\bibliography{ref.bib}
\end{document}